\documentstyle[12pt]{article}
\def\e{\mbox{e}}

\begin{document}
\title{Classical Preheating and Decoherence}
\author{D.~T.~Son\\
{\small\em  Department of Physics FM--15, University of Washington}\\
{\small\em Seattle, WA 98105, USA}}
\date{January 1996}
\maketitle
\begin{abstract}
We establish the equivalence between the quantum evolution of
spatially homogeneous oscillations of a scalar field and that of an
analogous classical system with certain random initial condition.  We
argue that this observation can be used for numerical simulation of
the Universe in the preheating epoch.  We also explicitly demonstrate
that the phenomenon of parametric resonance that leads to preheating
is simultaneously an effective mechanism for generating quantum
decoherence of the Universe.
\end{abstract}
\vskip 1in
UW/PT-96-01
\newpage

{\bf 1.} Calculation of the reheating temperature is one the most
important issue in inflationary cosmology.  The temperature achieved
soon after the inflation epoch is the maximal temperature in the whole
history of the Universe and must be sufficiently large for the
generation of baryon number to be effective.  According to the
inflationary picture, the Universe is reheated by the process of
transferring energy out of spatially homogeneous oscillations of the
scalar inflaton field.  The old theory of the reheating \cite{old}
describes this energy transfer as occurring through the decay of the
inflaton quanta and predicts a relatively low speed for this process
in theories with small couple constants.  However, a recent revision
\cite{KLS} shows that in many case there exists a more rapid channel
of energy transfer, which is closely related to the existence of
parametric resonance in the spectra of the fields interacting with the
inflaton and of the inflaton itself.  These parametric resonance
modes, typically forming continuous energetic bands, grow
exponentially on the background of the inflaton field oscillations,
and at some moment begin to take a large amount of energy from these
oscillations.  It has been argued in ref.\cite{KLS} that due to the
exponential character of the growth, the decay of the inflaton field
oscillations is a rapid process of ``explosion'' rather than a slow
damping.  The amplitude of these oscillations drops to a small value
after some time interval of the same order as of magnitude of the
period of oscillations.  This new mechanism, named in ref.\cite{KLS}
``preheating'', may change considerably the prediction of the
reheating temperature.  It has been argued that the phenomenon of
parametric resonance may be important in other aspects as well
\cite{KLS2,Tkachev}.

The complete theoretical description of preheating is still lacking
right now, due to the complexity of interaction between the modes that
appear via parametric resonance and the background, as well as between
the modes themselves.  However, some important steps have been made in
this direction (see, for example \cite{Shtanov,Boyanovsky}).  In one
of the approaches \cite{Boyanovsky} the problem is treated in the
Hartree--Fock approximation, where one takes into account the
back--reaction of the parametric--resonance modes to the background
but neglects the scattering between these modes.  The result is quite
surprising: while it confirms that parametric resonance is the main
mechanism of the damping of the inflaton field in the first time
period, in the long run the amplitude of the oscillations typically
remains constant (in some cases it is not much smaller than the initial
amplitude before the explosion), or at least decays very slowly.  This
shows that it takes much longer time, of many order of magnitudes
larger than the period of oscillations, to make the amplitude goes to
0.  At late times, the created particles are mostly of very small
energy, which means that the thermalization process may also take a
longer time interval than has been expected.  The calculations in
ref.\cite{Boyanovsky}, however, rely heavily on the Hartree--Fock
approximation, so there remains the question whether these results
reflect the real situation.

In this paper we do not attempt to build a complete theory of
preheating.  However, we will point out a possible way to perform
reliable calculations in the nonlinear regime after inflation.
Namely, we show that the behavior of the quantum system under
consideration, during the preheating epoch, is equivalent to the
evolution of the classical field, when one randomizes over a
particular set of initial conditions.  This result is established by
comparing the perturbative series of the two theories.  This fact
makes possible numerical simulations of the Universe after inflation,
since the problem is now completely classical.  As a by--product, we
demonstrate explicitly the development of decoherence of the Universe.
We show that the phenomenon of parametric resonance provides an
effective mechanism for the latter.

{\bf 2.} For simplicity we will work in the model of one scalar field
only. In other words, we neglect the interaction of the inflaton with
other particles and take into account its self--interaction only.  The
Lagrangian of the model is taken in the standard form,
\[
  L={1\over2}(\partial_{\mu}\phi)^2-{m^2\over2}\phi^2-
  {\lambda\over4!}\phi^4
\]
We will also neglect the expansion of the Universe and work in the
ordinary Minkowskian space--time.  At the moment when inflation
completes, the inflaton field has a nonzero expectation value
$\phi=\phi_0$ which is a constant over the whole space.  Soon after
that the field begins oscillations around $\phi=0$.  For convenience
the following picture is used in further discussions: at $t<0$ there
is an extra force term $J\phi$ in the Lagrangian so that the
expectation value of $\phi$ is a nonzero constant $\phi_0$, but at
$t=0$ this source is suddenly turned off and the field starts
oscillate.  We will be interested in the time dependence of the mean
value of $\phi$, which will be denoted as $\phi(t)$.

The linearized classical equation for the modes of momentum ${\bf k}$
of the field, $\phi(t,{\bf k})$,
\begin{equation}
  \left(\partial_t^2+\omega^2_k+{\lambda\over2}\phi^2_0(t)\right)
  \phi(t,{\bf k})=0
  \label{linear}
\end{equation}
has the same form as that of the oscillator whose frequency is a
periodic function of time,
$\omega^2=\omega_k^2+{\lambda\phi_0^2(t)\over2}$.  It is well known
that at some values of $\omega_k$ this system exhibits parametric
resonance: there does not exist any solution to eq.(\ref{linear})
which remains finite on the whole time axis \cite{Landau}.  A typical
solution to eq.(\ref{linear}) grows exponentially in the limits
$t\to+\infty$ and/or $t\to-\infty$.  This phenomenon occurs when
$\omega_k$ lies in certain energetic bands (``resonance bands'').  We
will consider the situation of broad resonance, which correspond to
large values of $\phi_0$, $\phi_0\sim1/\sqrt{\lambda}$ or larger.

In the quantum case, the existence of parametric resonance leads to
the amplification of the quantum fluctuations of the modes with
$\omega_k$ inside the resonance bands.  When the amplitude of these
fluctuations are still small, one can make use of the standard
Bogoliubov transformations to find the quantum state of the system
\cite{Shtanov}.  However, the interesting regime is that of late times
where the non--linearity becomes essential, and one needs another
approach to attack the problem.

The general method to find real--time evolution of quantum fields is
the Schwinger--Keldysh closed--time--path formalism
\cite{Landau2,Calzetta,Paz}.  At diagrammatic level, the formalism
gives rise to the same set of Feynman rules as that for calculating
the $S$--matrix, except that every internal integral in Feynman
diagrams is now performed along a contour that goes from $t=-\infty$
to $t=+\infty$ and then goes back to $t=-\infty$ (Fig.1), and instead
of one Feynman propagators $G(x,y)$ there exists four ones depending
on whether $x_0$ and $y_0$ lie on the upper or the lower parts of the
contours,
\[
  G_{++}(x,y)=\langle0|T\phi(x)\phi(y)|0\rangle,\qquad
  G_{+-}(x,y)=\langle0|\phi(y)\phi(x)|0\rangle,
\]
\[
  G_{-+}(x,y)=\langle0|\phi(x)\phi(y)|0\rangle,\qquad
  G_{--}(x,y)=\langle0|\bar{T}\phi(x)\phi(y)|0\rangle,
\]
It is easy to see that these propagators arise from the propagators on
the contour $C$, $G_C(x,y)=\langle0|T_C\phi(x)\phi(y)|0\rangle$, where
$T_C$ is the notation for time--ordering along $C$.  Note also that
the four propagators are equal at $x=y$.

To derive the Feynman rules, we will decompose the quantum field into
a classical part $\phi_0(t)$ and quantum fluctuation $\tilde{\phi}$,
\[
  \phi(t)=\phi_0(t)+\tilde{\phi}(t)
\]
where $\phi_0(t)=\phi_0$ at $t<0$ and at $t>0$ satisfies the field
equation
\[
  \partial_\mu^2\phi_0+m^2_0\phi_0+\lambda\phi_0^3=0
\]
The Lagrangian for $\tilde{\phi}$ is then
\[
  L(\tilde{\phi})={1\over2}(\partial_\mu\phi)^2-{m^2\over2}\tilde{\phi}^2
  -{\lambda\over4}\phi_0^2\tilde{\phi}^2-
  {\lambda\over3!}\phi_0\tilde{\phi}^3-{\lambda\over4!}\tilde{\phi}^4
\]
and corresponds to the Feynman rules shown in Fig.2.

Before turning to the calculation of diagrams, let us consider in more
details the propagators on the background $\phi_0(t)$.  Consider, for
instance, $G_{++}(x,y)$.  It is convenient to use the mixed
representation (coordinate in time and momentum in space) where this
propagators can be written into the form,
\[
  G_{++}(x,y)=\int\!{d{\bf k}\over(2\pi)^3}\,\e^{i{\bf k}({\bf x-y})}
  G_{++}(x_0,y_0;{\bf k})
\]
where
\begin{equation}
  iG_{++}(x_0,y_0;{\bf k})=
  \theta(x_0-y_0)f^k_1(x_0)f^k_2(y_0)+\theta(y_0-x_0)f^k_2(x_0)f^k_1(y_0)
  \label{G++}
\end{equation}
We have introduced the mode functions $f^k_{1,2}$ which are the two
linearly independent solutions to the equation
\begin{equation}
  \left(\partial^2_t+\omega_k^2+{\lambda\over2}\phi_0^2(t)\right)
  f_{1,2}^k(t)=0
  \label{modes}
\end{equation}
with the following boundary conditions at $t<0$,
\[
  f^k_1(t)\sim\e^{-i\omega_kt},\qquad f^k_2(t)\sim\e^{i\omega_kt}
\]
where $\omega_k=\sqrt{k^2+m^2+{\lambda\phi_0^2\over2}}$.  The mode
functions are normalized so that $f_1'(t)f_2(t)-f_1(t)f_2'(t)=-i$.  If
$\phi_0(t)$ is 0, $f^k_1$ and $f^k_2$ would be equal to
$\e^{-i\omega_kt}/\sqrt{2\omega_k}$ and
$\e^{i\omega_kt}/\sqrt{2\omega_k}$, respectively.  It is easy to see
that $f^k_2(t)=(f^k_1(t))^*$.

The formulas for other propagators are similar,
\[
  iG_{+-}(x_0,y_0;{\bf k})=f_2^k(x_0)f_1^k(y_0)
\]
\[
  iG_{-+}(x_0,y_0;{\bf k})=f_1^k(x_0)f_2^k(y_0)
\]
\begin{equation}
  iG_{--}(x_0,y_0;{\bf k})=
  \theta(x_0-y_0)f^k_2(x_0)f^k_1(y_0)+\theta(y_0-x_0)f^k_1(x_0)f^k_2(y_0)
  \label{G}
\end{equation}

Consider now a value of $k$ where there is parametric resonance in
eq.(\ref{modes}).  When parametric resonance is present,
eq.(\ref{modes}) possesses two real solutions $f_+^k(t)$ and
$f_-^k(t)$ which satisfy the following conditions at $t>0$
\cite{Landau},
\[
  f_+^k(t+T)=\e^{\epsilon_k T}f_+^k(t)
\]
\[
  f_-^k(t+T)=\e^{-\epsilon_k T}f_-^k(t)
\]
where $\epsilon_k$ is some real, positive constant depending on
$\omega_k$.  In particular, $f_+$ grows as $t\to+\infty$ while the
behavior of $f_-$ is opposite.  The previously defined mode functions
$f_1$ and $f_2$ are linear combinations of $f_+$ and $f_-$.  Not
interested in the particular form of the coefficients, we write,
\begin{equation}
  f_1^k(t)=\alpha_kf_+^k(t)+\beta_kf_-^k(t),\qquad 
  f_2^k(t)=\alpha_k^*f_+^k(t)+\beta^*_kf_-^k(t),
  \label{lincom}
\end{equation}
where we have made use of the fact that $f_1$ and $f_2$ are complex
conjugate and $f_+$ and $f_-$ are real.  Since as $f_+$ is growing
with $t$ while $f_-$ is falling, at large $t$ the part $f_+(t)$
dominates over $f_-(t)$.  If in eq.(\ref{lincom}) one neglects $f_-$
and leaves only $f_+$, there is no difference between $f_1$ and $f_2$,
except for an overall factor.  If one also neglects modes outside
resonance bands, which is natural since these modes are not enhanced,
and substituting eq.(\ref{lincom}) to eqs.(\ref{G++}), (\ref{G}) one
finds that the four propagators are equal to each other to the leading
order,
\[
  iG_{++}(x,y)=iG_{+-}(x,y)=iG_{-+}(x,y)=iG_{--}(x,y)=
\]
\begin{equation}
  =iG^0(x,y)=
  \int\!{d{\bf k}\over(2\pi)^3}\,|\alpha_k|^2f_+^k(x_0)f_+^k(y_0)
  \label{6}
\end{equation}

Now let us consider the one--loop contribution to $\phi$.  The only
diagram that contributes to $\phi$ is the tadpole one shown in Fig.3a.
This diagram is equal to
\begin{equation}
  \phi_1(x)={i\lambda\over2}\int\limits_C\!dy_0
   \!\int\!d{\bf y}\,G_C(x,y)\phi_0(y_0)G(y,y)
  ={i\lambda\over2}\int\!dy\,G_R(x,y)\phi_0(y_0)G(y,y)
  \label{GRG}
\end{equation}
where we have dropped the indices of $G(y,y)$ since all the four Green
functions are equal at coinciding points and introduced the retarded
Green function, $G_R=G_{++}-G_{+-}$.  One can expect that at late
$x_0$ the integral in eq.(\ref{GRG}) is dominated by large values of
$y_0$ where $G(y,y)$ can be replaced by the leading term,
eq.(\ref{6}).  So one finds,
\begin{equation}
  \phi_1(x_0)={\lambda\over2}\int\!dy_0\,G_R(x_0,y_0;0)
  \phi_0(y_0)\int\!{d{\bf k}\over(2\pi)^3}\,|\alpha_kf^k_+(y_0)|^2
  \label{1loop}
\end{equation}
Eq.(\ref{1loop}) can be represented graphically as in Fig.3b, where
each external line ending with a bullet is associated with the factor
$|\alpha_k|f^k_+(y_0)$ if ${\bf k}$ is a resonance mode and 0 in the
opposite case.  Note that $f^k_+$ grows exponentially with $t$, so
$\phi_1$ is also growing.  At $t$ when $\phi_1$ becomes comparable to
$\phi_0$ one can expect that all terms of the perturbative series are
of the same order, and the perturbation theory breaks down.  We will
try to extract the main contribution from each order of the
perturbation theory in this regime.  We will not be interested at
latter times when $\phi_1\gg\phi_0$.

Let us move to two--loop diagrams.  Consider, for example, the one
depicted in Fig.4a,
\begin{equation}
  \phi_2(x_0)={i\lambda^2\over2}\int\limits_C\!dy_0\,dz_0\,
  {d{\bf k}\over(2\pi)^3}\,G_C(x_0,y_0;0)\phi_0(y_0)
  G_C(y_0,z_0;{\bf k})G_C(y_0,z_0;{\bf k})\phi_0(z_0)\phi_1(z_0)
  \label{2loop}
\end{equation}
If one is interested large values of $x_0$, it can be expected that
the important region of integration in r.h.s. of eq.(\ref{2loop}) is
that of large $y_0$ and $z_0$.  So, one may suspect that the integral
can be calculated by replacing $G_C(x_0,y_0;{\bf k})$ by its leading
order contribution, eq.(\ref{6}).  One obtains after that
\[
  \phi_2(x_0)=-{i\lambda^2\over2}\int\limits_C\!dy_0\,dz_0\,  
  {d{\bf k}\over(2\pi)^3}\,G_C(x_0,y_0;0)\phi_0(y_0)
  (f^k_+(y_0))^2(f^k_+(z_0))^2\phi_0(z_0)\phi_1(z_0)
\]
However, it is easy to see that the integration over $z_0$ gives zero,
since it goes along the contour $C$ and the integrand,
$(f^k_+(z_0))^2\phi_0(z_0)\phi_1(z_0)$, is the same on the upper and
lower parts of the contour.  So, one should turn to the
next--to--leading order.  Denoting the next correction to eq.(\ref{6})
as $G^1_C(x,y)=G_C(x,y)-G^0(x,y)$, one writes,
\[
  \phi_2(x_0)=\lambda^2\!\int\limits_C\!dy_0\,dz_0\,
  {d{\bf k}\over(2\pi)^3}\,G_C(x_0,y_0;0)\phi_0(y_0)
  G^1_C(y_0,z_0;{\bf k})f_+^k(y_0)f_+^k(y_0)\phi_0(z_0)\phi_1(z_0)
\]
\begin{equation}
  =\lambda^2\!\int\limits_C\!dy_0\,dz_0\,
  {d{\bf k}\over(2\pi)^3}\,G_C(x_0,y_0;0)\phi_0(y_0)
  G_C(y_0,z_0;{\bf k})f_+^k(y_0)f_+^k(y_0)\phi_0(z_0)\phi_1(z_0)
  \label{2looptree}
\end{equation}
where in the last equation we have replaced $G^1_C$ by $G_C$, making
use of the fact that the leading term vanishes.  Eq.(\ref{2looptree})
can be rewritten in a simpler form using the retarded propagator
$G_R=G_{++}-G_{+-}$,
\[
  \phi_2(x_0)=\lambda^2\!\int\limits_{-\infty}^{\infty}\!dy_0\,dz_0\,
  {d{\bf k}\over(2\pi)^3}\,G_R(x_0,y_0;0)\phi_0(y_0)
  G_R(y_0,z_0;{\bf k})f_+^k(y_0)f_+^k(y_0)\phi_0(z_0)\phi_1(z_0)
  \label{2looptree'}
\]
where the integrations over $dy_0$ and $dz_0$ are performed from
$-\infty$ to $+\infty$.  Eq.(\ref{2looptree}) can be represented
graphically as in Fig.4b.  As in the case of the one--loop graph, the
calculation of the loop diagram, thus, is reduced to evaluation of a
tree connected graph.

This technique can be generalized to deal with other diagrams and
higher orders of perturbation theory.  Summarily, the technique is the
following.  To calculate a given Feynman diagram with $l$ loops, one
cuts $l$ internal propagators, so that the obtained diagram is tree,
and connected.  If there are many ways to perform this cutting, one
should take a sum over all possibilities.  Each propagator that has
been cut is associated with two factors of $f_+^k$ attached to the two
vertices that the former propagator had been connecting.  The
propagators that have not been cut are associated with the retarded
propagator $G_R$.  The obtained diagrams are computed by the standard
way, where the integral are performed along the usual time axis.  We
note that this reduction of loop graphs to tree ones is very similar
to that found in another setting \cite{LRST}.  We emphasize that our
method is applicable in the regime when all terms of the perturbation
series are of the same order.  At latter times when $\phi_1\gg\phi_0$
our approach does not work, since the pieces that have been neglected
may become important.

{\bf 3.} The fact that the diagrams obtained after the cutting
procedure are tree points to the possibility of classical description
of the problem.  To find the latter, let us consider the Cauchy
initial problem with the classical field equation
\begin{equation}
  (\partial_{\mu}^2+m^2)\phi+\lambda\phi^3=0
  \label{background}
\end{equation}
with the following initial condition at the asymptotics $t\to-\infty$,
\begin{equation}
  \phi(t,{\bf x})=\phi_0(t)+\int\!{d{\bf k}\over(2\pi)^3}\,
  c_{\bf k}f_+^k(t)\e^{i{\bf kx}}
  \label{linear_bc}
\end{equation}
where $c_{\bf k}$ are some arbitrary set of complex numbers.  For
$\phi$ to be real, we require that $c_{\bf k}^*=c_{\bf -k}$.  Since at
$t\to-\infty$ $f^k_+(t)$ is small and satisfies eq.(\ref{modes}),
eq.(\ref{linear_bc}) is the solution to the field equation up to the
first order of $f_+$.  The corrections to eq.(\ref{linear_bc}), which
start at the second order of $f_+$, can be calculated iteratively from
the field equation and the result can be represented in the form of
tree Feynman diagrams.  For example, the second--order contribution is
represented by the first diagram in Fig.5 together some third-- and
fourth--order graphs.  In these graphs each external leg ending on a
bullet is associated with the factor $c_{\bf k}f_+^k$, and each
internal line corresponds to the retarded propagator.

Despite the fact that these diagrams are similar to that obtained by
the cutting procedure described above, there are two main differences
between the two cases.  First, the diagrams obtained by cutting loop
graphs always contains an {\em even} number of external bullet legs.
Second, the external bullet legs in the first case can be grouped into
pairs so that the momenta running along legs of the same pair are of
opposite signs, while in the graphs coming from eq.(\ref{linear_bc})
the only requirement is that the sum of the momenta of the bullet legs
is 0.  Nevertheless, it is easy to show that there is a simple
relation between the two cases.  Namely, if one denotes as
$\phi[t,{\bf x},c_{\bf k}]$ the solution to the field equation with
the boundary condition (\ref{linear_bc}), then the quantum average of
$\phi$ in our problem is given by an integral over $c_{\bf k}$ with a
Gaussian measure,
\begin{equation}
  \phi(t)=\int\!{\cal D}c_{\bf k}
  \exp\left(-\int\!{d{\bf k}\over(2\pi)^3}\,
  {|c_{\bf k}|^2\over2|\alpha_k|^2}\right)
  \phi[t,{\bf x},c_{\bf k}]
  \label{average}
\end{equation}
(the integral in the r.h.s.\ of eq.(\ref{average}) does not depend on
${\bf x}$).  To see this, one notices that the integration over ${\cal
D}c_k$ leaves only diagrams with even number of bullet legs which can
be grouped in paired of opposite momenta.  The integration also adds a
factor of $|\alpha_k|^2$ to each pair with momentum $({\bf k},{\bf
-k})$.  So, the graphical representation of the r.h.s.\ of
eq.(\ref{average}) is the same as those obtained by cutting the loop
graphs.

In eq.(\ref{average}), $c_{\bf k}$ are supposed to be real,
Gaussian--distributed around 0 with the standard deviation
$|\alpha_k|$ and the $c_{\bf k}$ with different ${\bf k}$ are not
correlated.  Eq.(\ref{average}) is not the only possible, one could
write, for example
\begin{equation}
  \phi(t)=\int\!{\cal D}\theta_{\bf k}\,
  \phi[t,{\bf x},|\alpha_k|\exp(i\theta_{\bf k})]
  \label{average'}
\end{equation}
where $c_{\bf k}$ now are complex numbers with fixed absolute value
equal to $|\alpha_k|$ but with random phase.  One can write more
relations of this type.

Eqs.(\ref{average},\ref{average'}) give rise to the possibility of
numerical simulation of the processes occurring in the Universe after
inflation.  To do this it is sufficient to take an initial field
configuration in the form of a homogeneous oscillations plus
appropriate random perturbations, and evolve this configuration in
time according to the field equation.  The mean value of $\phi$ in
this classical system is the same as that of the quantum one.  It is
trivial to derive the same result in theories with more than one
scalar fields.

Eqs.(\ref{average},\ref{average'}) are the manifestation of the fact
that when the amplitude of quantum fluctuations is large (roughly
speaking, when the occupation numbers are large), the system behaves
like a classical one.  Our derivation gives this statement a precise
meaning.

It is also interesting to mention the relation between
eq.(\ref{average},\ref{average'}) and the fundamental phenomenon of
{\em decoherence} of the Universe.  In a situation when, as in our
case, a closed system starts its evolution from a quantum pure state
at $t=0$ it will remains in a pure state at any time moment.  The
system that we are considering can be described, in the pure state
language, by the operators obtained by Bogoliubov transformation
\cite{Shtanov}.  However, when the resonance modes become large the
modes strongly interact between themselves and the pure state
description becomes very complicated.  In rescue, another description
comes into effect: the system can be considered as a mixed state,
i.e. that of an ensemble, of states (semiclassical in our case)
characterized by the solution to the field equation with initial
condition (\ref{linear_bc}) with $c_{\bf k}$ playing the role of the
indices numbering the quantum states in the ensemble.  Explicitly, the
system is described by the following density matrix,
\[
  \hat{\rho}\sim\sum_{\{c_k\}}|\phi[t,{\bf x},c_{\bf k}]\rangle
  \rho[c_{\bf k}]\langle\phi[t,{\bf x},c_{\bf k}]|
\]
where $|\phi[t,{\bf x},c_{\bf k}]\rangle$ is the notation for a (pure)
semiclassical quantum state associated with the classical field
configuration $\phi[t,{\bf x},c_{\bf k}]$ (the choice of $|\phi[t,{\bf
x},c_{\bf k}]\rangle$ is not unique).
Eq.(\ref{average},\ref{average'}) ensures that the average of $\phi$
in this mixed state is equal to its true quantum average in the pure
state with proper choice of $\rho[c_{\bf k}]$.  It can be shown that
the same is valid for other Green functions as well.  So, for
calculating physical quantities one can consider the system as being
in a mixed state, though in fact its true state is a pure one.  Note
that the time period when $c_{\bf k}$ are large, $c_{\bf k}\gg1$, but
still sufficiently small so that the modes are linear, is the matching
region where both pure-- and mixed--state descriptions can be used.

{\bf 4.} So, we have shown that the problem of evolution of the
 homogeneous oscillating scalar field that has a direct
connection with preheating in inflationary cosmology is equivalent
to the classical evolution of the scalar field with some random
initial conditions.  We also show that the parametric resonance
phenomenon provides an effective mechanism for the generation of
decoherence of the Universe.  We have seen a nice feature of this
mechanism that the transition from the pure--state to mixed--state
description can be traced explicitly.  

The result of this paper may be useful for numerical modeling of the
preheating process and hopefully will leads to the better
understanding of the latter.  Preliminary results seems to agree
qualitatively with that of the the Hartree--Fock approximation: the
fall of the amplitude of oscillations is saturated at some finite
level.  Moreover, after this saturation the modes with small ${\bf k}$
seem to play an important role in the evolution.  The work is still
continuing, results will be published elsewhere.

The author thanks P.~Arnold, V.~Rubakov, P.~Tinyakov and L.~Yaffe for
valuable discussions.

\newpage

\setlength{\unitlength}{1cm}

\begin{figure}
\begin{center}
\begin{picture}(7,2)
\put(0,1){\vector(1,0){7}}
\put(2,0){\line(0,1){2}}
\put(0,1.25){\vector(1,0){3}}
\put(3,1.25){\line(1,0){3}}
\put(6,0.75){\vector(-1,0){3}}
\put(3,0.75){\line(-1,0){3}}
\put(6,1){\oval(0.5,0.5)[r]}
\put(6.8,0.7){$t$}
\put(0,1.3){$C$}
\end{picture}
\end{center}
\caption{Integration contour in Schwinger--Keldysh formalism.}
\label{1}
\end{figure}
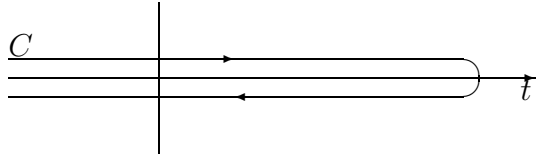

\setlength{\unitlength}{1pt}
\begin{figure}
\thicklines
\begin{center}
\begin{picture}(380,120)(30,240)
\put(100,250){\line(1,0){50}}
\put(200,246){$i(-\partial^2+m^2+{\lambda\over2}\phi_0^2)^{-1}$}
\put(50,330){\line(1,0){30}}
\put(50,330){\line(-3,-5){20}}
\put(50,330){\line(-3,5){20}}
\put(130,326){$\displaystyle{-i\lambda\phi_0}$}
\put(310,330){\line(1,1){25}}
\put(310,330){\line(1,-1){25}}
\put(310,330){\line(-1,1){25}}
\put(310,330){\line(-1,-1){25}}
\put(385,326){$\displaystyle{-i\lambda}$}
\end{picture}
\end{center}
\caption{Feynman rules for calculating $\phi(t)$.}
\label{feyn_rules}
\end{figure}
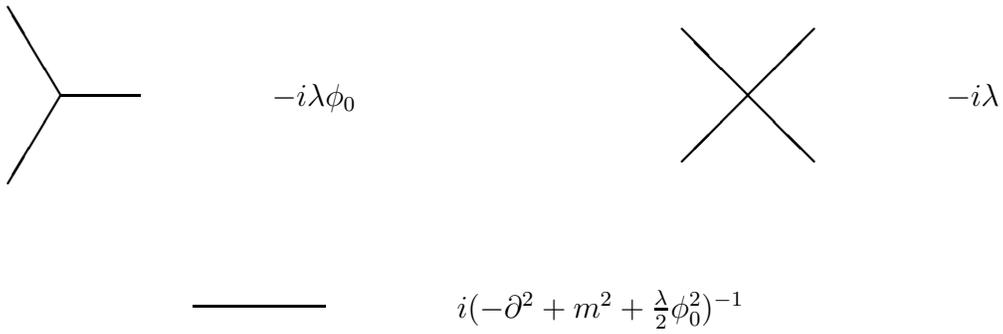

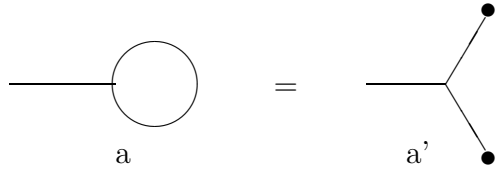
\begin{figure}
\begin{center}
\begin{picture}(185,60)(105,300)
\put(105,330){\line(1,0){40}}
\put(160,330){\circle{30}}
\put(205,326){$\displaystyle{=}$}
\put(145,300){a}

\put(240,330){\line(1,0){30}}
\put(270,330){\line(3,5){15}}
\put(270,330){\line(3,-5){15}}
\put(286,358){\circle*{5}}
\put(286,302){\circle*{5}}
\put(255,300){a'}
\end{picture}
\end{center}
\caption{The one--loop graph (a) and the representation (b) of its leading
contribution.}
\label{fig:1loop}
\end{figure}

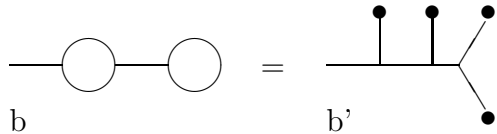
\begin{figure}
\begin{center}
\begin{picture}(185,50)(15,225)

\put(15,250){\line(1,0){20}}
\put(45,250){\circle{20}}
\put(55,250){\line(1,0){20}}
\put(85,250){\circle{20}}
\put(110,246){$\displaystyle{=}$}
\put(15,225){b}

\put(135,250){\line(1,0){50}}
\put(185,250){\line(3,5){10}}
\put(185,250){\line(3,-5){10}}
\put(196,270){\circle*{5}}
\put(196,230){\circle*{5}}
\put(155,250){\line(0,1){20}}
\put(175,250){\line(0,1){20}}
\put(155,270){\circle*{5}}
\put(175,270){\circle*{5}}
\put(135,225){b'}
\end{picture}
\end{center}
\caption{A two--loop graph (a) and its reduction to a tree graph (b).}
\label{fig:2loop}
\end{figure}

\begin{figure}
\begin{center}
\begin{picture}(330,105)(0,0)
\put(0,50){\line(1,0){50}}
\put(50,50){\line(1,1){25}}
\put(50,50){\line(1,-1){25}}
\put(75,75){\circle*{5}}
\put(75,25){\circle*{5}}

\put(100,50){\line(1,0){50}}
\put(150,50){\line(1,-1){25}}
\put(150,50){\line(1,1){25}}
\put(175,75){\line(1,0){25}}
\put(175,75){\line(0,1){25}}
\put(175,25){\circle*{5}}
\put(200,75){\circle*{5}}
\put(175,100){\circle*{5}}

\put(225,50){\line(1,0){50}}
\put(275,50){\line(1,-1){25}}
\put(275,50){\line(1,1){25}}
\put(300,25){\line(0,-1){25}}
\put(300,25){\line(1,0){25}}
\put(300,75){\line(0,1){25}}
\put(300,75){\line(1,0){25}}
\put(300,0){\circle*{5}}
\put(325,25){\circle*{5}}
\put(325,75){\circle*{5}}
\put(300,100){\circle*{5}}
\end{picture}
\end{center}
\caption{Some tree diagrams arising from the classical problem.}
\label{fig:classical}
\end{figure}
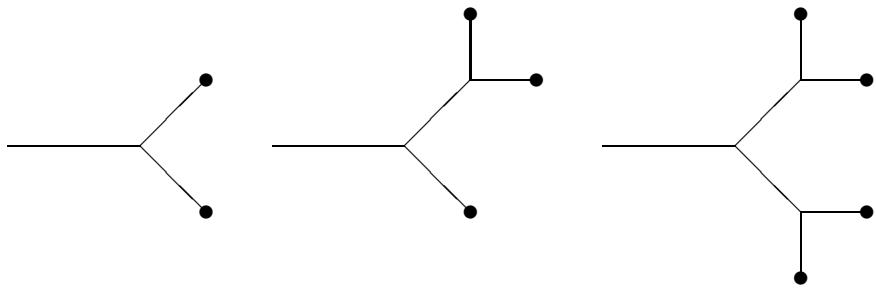

\end{document}